\documentstyle[12pt]{article}
\begin{document}

\begin{flushright}
KOBE--FHD--93--05\\
May  1994\\
\end{flushright}
\vspace{2em}

\begin{center}
\renewcommand{\thefootnote}{\fnsymbol{footnote}}
{\large \bf Planckian Scatterings of Massive Particles\\
            and Gravitational Shock Waves}\\
\vspace{4em}
Koichi HAYASHI${}^{1}$ and Toshiharu SAMURA${}^{2}$
          \footnote[2]{e-mail address: samura@cphys.cla.kobe-u.ac.jp}\\
\vspace{3em}
${}^{1}${\it Department of Mathematics and Physics,}\\
        {\it Faculty of Science and Technology,}\\
        {\it Kinki University, Higashi-Osaka,}\\
		{\it  Osaka 577, Japan}\\
\vspace{2em}
${}^{2}${\it Graduate School of Science and Technology, }\\
        {\it Kobe University, Nada, Kobe 657, Japan}\\
\end{center}
\vspace{2em}

\begin{center}
\begin{abstract}
\baselineskip 12pt
The scattering process of two particles at Planck energies or beyond
is calculated using the gravitational shock wave metric for a massive black
hole.
Then, the scattering between a heavy mass particle and a small mass one is
deal with.
The cross section contains an extra new term of $\sigma \propto G^3$
as the correction term of the leading term derived by 't Hooft.
The ultrarelativistic limit of the Lorentz boosted Reissner--Norstrom
black hole is also calculated.
\end {abstract}
\end{center}

\clearpage
\baselineskip 16pt

After the paper of 't Hooft \cite{'t Hooft87}, many articles have been devoted
to
the study of the particle scatterings at Planck scale.
When particle scatterings take place at energies of the order or larger than
Planck
mass, $M_{Planck}$, the gravitational interaction
dominates their collision processes. Therefore this would be useful to
the understanding of quantum gravity. Although a conventional quantum field
theory of the gravitation is unrenormalizable, physical insights at Planck
scale can
be obtained from the study of quantum fields in classical geometries.
For example, the classical background geometry of the ultrarelativistic
particle is chosen as the gravitational shock wave ( G. S. W.) metric of a
black
hole.

The G. S. W. metric of a black hole is the one which makes a black hole
move with an ultrarelativistic speed
( Lorentz $\gamma$ factor $\rightarrow \infty$).
As the black hole is Lorentz contracted, it looks like a plane black hole at
this limit.
In usual cases, G. S. W. metrics for the massless Schwarzschild black hole
is obtained by
Aichelburg and Sexl (AS) \cite{AS71}.
On the other hand, Loust\'o and S\'anchez derived the G. S. W. metric
for the massless Kerr--
Newman black hole \cite{LS92}
and the massless Reissner--Norstrom (RN) black hole\cite{LS90}.

Recently, the authors have derived the G. S. W. metric for a massive black hole
( the Schwarzschild and Kerr black hole) \cite{HS94} (HS below).
We have assumed that $\gamma$
factor becomes very large but finite. This is nothing but leading contribution
for
large $\gamma$ ( not mathematically infinite) in $1/\gamma$ expansion is
sought.
For a particle scattering at Planckian energy or beyond, the $\gamma$
factor is very large but finite. Therefore our derived metrics can be applied
for the scattering of the massive particles.
The mass of the black hole takes an arbitrary and finite value.

Using the method of Loust\'o and S\'anchez, the G. S. W. metric for a massive
Schwarzschild black hole becomes \cite{HS94}:
\begin{eqnarray}
&&\lefteqn{\lim_{\gamma \to \infty} ds^2_S \longrightarrow
    -du~dv+dx^2+dy^2}\nonumber \\
&&-4p \left[ \log \left( \frac{\rho}{2GM} \right)^2  -\frac{\rho}{4M}
   \left(\pi-4 \sqrt{ 1-\frac{4M^2}{\rho^2} }
    \tan^{-1}  \sqrt{\frac{\rho+2M}{\rho-2M}} \right) \right]
  \delta \left( u \right)  du^2,\nonumber\\
\end{eqnarray}
where a Lorentz boost in the $z$--direction is applied and we put
$u=t-z$, $v=t+z$. $\rho$ is the length perpendicular to the motion,
$M$ is the mass of the black hole, and $p=\gamma M$.
{}From this metric,
two properties of the geodesic, the shift, $\Delta v$, of a null coordinate $v$
at $u=0$,
and the refraction angle crossing the G. S. W. are obtained
\cite{AS71,'t Hooft85,HS94}.
Although the denominator of the logarithmic term is an irrelevant constant,
we normalized it at the horizon radius.
For $2M/\rho \ll 1$, Eq.(1) is reduced to:
\begin{eqnarray}
&&\lim_{\gamma \to \infty} ds^2_S {\longrightarrow \atop \rho \gg 2M}
    -du~dv+dx^2+dy^2\nonumber \\
&&-4p\left[ \log \left( \frac{\rho}{2GM} \right)^2  +1-\frac{\pi}{2}
    \frac{M}{\rho}
       +O\left( \frac{M^2}{\rho^2} \right) \right]
  \delta \left( u \right)  du^2~~,
\end{eqnarray}
and the shift $\Delta v$ is:
\begin{eqnarray}
\Delta v~=~-8p\left[ \log \left( \frac{\rho}{2GM} \right)^2
    +1-\frac{\pi}{2} \frac{M}{\rho}
        +O\left( \frac{M^2}{\rho^2} \right) \right]~~.
\end{eqnarray}

In this paper, we derive the G. S. W. metric for a massive RN
black hole. The extremal hole is given for the limit of $\sqrt{G} M \rightarrow
Q$
( $Q$ : the charge of the black hole).
In other words, $M \sim M_{Planck}$ for $Q \sim e$ ( $e$ : the electron charge)
so that it corresponds to the scattering of massive charged
particles at Planck energies.

In general, the G. S. W. metric is relevant to the scattering of particles at
high
energy. Point particle scattering \cite{'t Hooft87,point}
 and string scattering \cite{string} in the metric
for the massless Schwarzschild black hole, and for massless Kerr--Newman
black hole \cite{LS92}
and for cosmological defects \cite{LS91} have been investigated.
The importance of the black hole S matrix are also studied
\cite{'t Hooft90}.

't Hooft calculated the S matrix of the leading order ($\log \rho^2$ term)
 using AS metric \cite{'t Hooft87,'t Hooft90}.
Here, we study the particle scattering to the next order in $1/\rho$,
using Eq.(3).
It is shown that the correction term
of the leading term derived by 't Hooft can be calculated since we have not
assumed $M \rightarrow 0$.
So far, the correction terms to the leading term $\log \rho^2$ have
been calculated \cite{point,string}. But they are of order
$O \left( 1/\rho^2 \right)$,
because the limits $M \rightarrow 0$ are taken.
Since we have not assumed $M \rightarrow 0$, our modification appears
in $O \left(M/\rho \right)$, which is a new result.
The scattering cross section, $\sigma$, is obtained and it contains naturally
 the term of one graviton exchange
($\sigma \propto G^2$) and the two gravitons exchange
($\sigma \propto G^4$).
It is to be noted, however, that it contains an extra new term of
$\sigma \propto G^3$ as the interference between the two.

First, we calculate the G. S. W. metric of a RN
black hole by the method discussed in the latest paper \cite{HS94}.
We put $G=c=\hbar=1$ for a while.
A RN metric in the ordinary
coordinates is divided into the Minkowski metric $ds_M^2$ term and the
surplus term $\Delta S_{RN}^2$:
\begin{eqnarray}
ds^2_{RN} &=& -\left(1-2M/\bar{r}+Q^2/\bar{r}^2 \right) d\bar{t}^2
       +\left(1-2M/\bar{r}+Q^2/\bar{r}^2 \right)^{-1} d\bar{r}^2\nonumber\\
	   &&~~~+\bar{r}^2 d\bar{\theta}^2 +
          \bar{r}^2 \sin ^2\bar{\theta} d\bar{\phi}^2 \nonumber \\
     &=& ds^2_M +\Delta S^2_{RN}~~,
\end{eqnarray}
where
\begin{eqnarray}
ds_M^2 &=& -d\bar{t}^2 +d\bar{r}^2 +\bar{r}^2 d\bar{\theta}^2 +
                \bar{r}^2 \sin ^2\bar{\theta}~d\bar{\phi}^2~~, \nonumber\\
\Delta S^2_{RN} &=&
                  \left( 2M/\bar{r}+Q^2/\bar{r}^2 \right) d\bar{t}^2 +
                    \left( 2M/\bar{r}-Q^2 \right) /
					   \left(1-2M/\bar{r}+Q^2 \right) d \bar{r}^2 ~~.
\end{eqnarray}
We set $\bar{r}^2=\bar{x}^2+\bar{y}^2+\bar{z}^2$.
$M$ and $Q$ are the mass and the charge of the black hole,
respectively.
Now, a Lorentz boost with velocity $v$ is performed in the
$z$--direction:
\begin{eqnarray}
t=\gamma \left( \bar{t}+v \bar{z} \right)~~,~
z= \gamma \left(\bar{z}+v \bar{t} \right)~~,~
x=\bar{x}~~,~y=\bar{y}~~,
\end{eqnarray}
where
\begin{eqnarray}
\gamma~=~(1-v^2)^{-1/2} ~~.
\end{eqnarray}
As $ds_{M}^2$ is invariant for Lorentz transformations, only
$\Delta S_{RN}^2$ is transformed. When the black hole is moving
with the limit $v \rightarrow 1$, old coordinates, $\bar{x}^i_{s}$,
are changed as like Eq.(2.3) of HS \cite{HS94}.

Then, $\Delta S_{RN}^2$ of Eq.(5) is transformed as follows:
\begin{eqnarray}
\Delta S^2_{RN} &= &\left[ \frac{2p}{\sqrt{u^2 +\tilde{\rho}^2}}
            -\frac{Q^2}{u^2+\tilde{\rho}^2}  \right.\nonumber\\
            &&~~\left.+\frac{2p~u^2}
			{\left[u^2+\left(\tilde{\rho}^2+\tilde{Q}^2 \right)
			-2\tilde{M}
              \sqrt{ u^2+\tilde{\rho}^2}\right]
			  \sqrt{u^2+\tilde{\rho}^2} } \right.\nonumber\\
			 &&~~\left.-\frac{Q^2~u^2}{\left[u^2+\left(\tilde{\rho}^2
			 +\tilde{Q}^2 \right)
			 -2\tilde{M}
              \sqrt{ u^2+\tilde{\rho}^2}\right] \left(u^2+\tilde{\rho}^2\right)
}
			  \right] du^2~~,
\end{eqnarray}
where we set the energy $p=\gamma M$, $\tilde{M}=M/\gamma$
and $\tilde{Q}=Q/\gamma$.

When the limit $v \rightarrow 1$ is taken, we carry out by the method of
Loust\'o
and S\'anchez \cite{LS92,LS90,LS91} : integrate over the variable $u$,
then take the limit
$v \rightarrow 1$ and finally take the derivative with respective to $u$ and
remove the redundant terms.
Introducing a new coordinate
\begin{eqnarray}
dv'~=~dv-4p~du/|u|~~,
\end{eqnarray}
the G. S. W. metric has the final form ( dash on $v'$ is omitted),
\begin{eqnarray}
\lim_{\gamma \to \infty} ds^2_{RN}~\longrightarrow~&&-du~dv
        +dx^2+dy^2 \nonumber\\
&&-4p \left[ \Delta g_1+\Delta g_2+\Delta g_3\right]~\delta(u)~du^2~~,
\end{eqnarray}
where
\begin{eqnarray}
\Delta g_1~&&=~\log \rho^2-\frac{2}{\alpha-\beta}
       \left\{ \sqrt{\alpha}
  \left( 1+\beta \right)
  \tan^{-1} \left( 1/\sqrt{\alpha} \right)\right. \nonumber\\
 &&\left. -\sqrt{\beta} \left( 1+\alpha \right)
	 \tan^{-1} \left( 1/\sqrt{\beta} \right) \right)~~, \nonumber\\
\Delta g_2~&&=~\frac{2 \rho Q^2}
                {M \left\{\rho \left(\sqrt{M^2-Q^2}+M \right) +Q^2 \right\}
}~~,
				\nonumber\\
 &&~~~~\times~\left\{ \frac{1}{\alpha-1}
  \left[ \frac{\pi}{4}-\frac{1}{\sqrt{\alpha}}
           \tan^{-1} \left( 1/\sqrt{\alpha} \right) \right] \right. \nonumber\\
 &&~~~~ \left. +\frac{\beta}{\beta-\alpha}
  \left[ \frac{1}{\sqrt{\beta}} \tan^{-1} \left( 1/\sqrt{\beta} \right)
     -\frac{1}{\sqrt{\alpha}} \tan^{-1} \left( 1/\sqrt{\alpha} \right) \right]
  \right\}~~, \nonumber\\
\Delta g_3~&&=~\frac{\pi~Q^2}{4 \rho M}~~, \nonumber\\
\end{eqnarray}
and
\begin{eqnarray}
\alpha~&=&~\frac{\rho^2-Q^2+2 \rho \sqrt{M^2-Q^2}}
                             {\rho^2+Q^2+2M \rho}~~, \nonumber\\
\beta~&=&~\frac{\rho^2-Q^2-2 \rho \sqrt{M^2-Q^2}}
							{\rho^2+Q^2+2M \rho}~~,
\end{eqnarray}
This is the exact result.

Next, we deal with the case of the extremal ($M \rightarrow Q$) RN black hole.
If $Q$ is equal to $e$, the electronic charge, $M \sim M_{Planck}$
in this limit.
It corresponds to the scattering of charged particles of mass $\sim
M_{Planck}$.
In this limit, Eqs.(10) has the following form:
\begin{eqnarray}
&&\lim_{M \longrightarrow Q}
     \left[ \lim_{\gamma \longrightarrow \infty} ds^2_{RN} \right]
 ~\longrightarrow ~-du~dv+dx^2+dy^2 \nonumber\\
&&~-4p \left[ \log \rho^2 +\frac{1}{2}
     +\frac{\rho^2-2Q^2}{Q \sqrt{\rho^2-Q^2}}
   \tan^{-1} \sqrt{\frac{\rho+Q}{\rho-Q}}
       -\frac{\rho^2-Q^2}{4\rho Q} \pi
 \right] \delta \left( u \right)~ du^2~~. \nonumber\\
\end{eqnarray}
The refraction angle becomes $90^\circ$ for $\rho =Q$, the event horizon.
It is realized that the event horizon is invariant under the Lorentz
transformation
for the $z$--direction because the event horizon of the extremal RN black hole
is at $\rho =Q$.

Finally, when $\rho \gg Q$, Eq.(13) is reduced to
\begin{eqnarray}
&&\lim_{M \longrightarrow Q}
     \left[ \lim_{\gamma \longrightarrow \infty} ds^2_{RN} \right]
	  {\longrightarrow \atop \rho \gg Q} -du~dv+dx^2+dy^2 \nonumber\\
   &&~-4p \left[ \log \rho^2+1-\frac{\pi}{8} \left( \frac{Q}{\rho} \right)
  -\frac{2}{3} \left( \frac{Q^2}{\rho^2} \right)
     +O\left( \frac{Q^3}{\rho^3} \right) \right]
     \delta \left( u \right)~du^2.
\end{eqnarray}
{}From Eq.(1.3) of HS \cite{HS94}, the shift $\Delta v$ in crossing $u=0$ is
\begin{eqnarray}
   \Delta v~=~-8p \left[ \log \left( \frac{\rho}{Q} \right)^2+1-\frac{\pi}{8}
   \left( \frac{Q}{\rho} \right)
  -\frac{2}{3} \left( \frac{Q^2}{\rho^2} \right)
     +O\left( \frac{Q^3}{\rho^3} \right) \right]
     \delta \left( u \right)~du^2.
\end{eqnarray}
The denominator in the logarithm is an irrelevant constant, but we
normalized it at the event horizon.

Next, consider the scatterings of two particles with masses $m^{(1)}$ and
$m^{(2)}$
, where $M_{Planck} \geq m^{(2)} \gg m^{(1)}$.
Here, we put $c=\hbar=1$ again, but the gravitational constant $G$ is
recovered.
The particle (1) is at rest
or moves slowly, whereas the particle (2) is moving with $v \sim 1$.

The initial state of the gravitational scattering is defined by C.M. energy $E$
larger than $M_{Planck}$ with the impact parameter $b$
conjugate to the momentum transfer $Q \ll E$.
In this paper, as a first step, the gravitational scattering to the
second--order
eikonal approximation is calculated.
We assume the following restriction.
\begin{eqnarray}
 E~\gg~M_{Planck},~~b~\gg~Gm^{(i)}~~.
\end{eqnarray}
$Gm^{(i)}$ is the horizon radius of the particle ($i$).

First, we consider the case of two electrically neutral particles.
The energy is so tremendous that we can no longer ignore the gravitational
field
of the particle (2). This field is represented by the G. S. W. metric of
Eq.(3).
According to the notation of 't Hooft (Eq.(7) of ref.\cite{'t Hooft87}),
the shifted wave function is given by:
\begin{eqnarray}
\Psi^{(1)}_+~=~\int A \left( k_+, \tilde{k} \right) dk_+~d^2 \tilde{k}
      \exp \left( i\tilde{k} \tilde{x} -i k_+ u-ik_{-} v \right)~~,
\end{eqnarray}
where
\begin{eqnarray}
&&A \left( k_+, \tilde{k} \right)~=~
        \delta \left( k_+ -p_+^{(1)} \right) \nonumber\\
		&&~~\times \frac{1}{\left(2 \pi \right)^2} \int d^2 \tilde{x}
		    ~\exp \left\{ i\tilde{k} \tilde{x}
			    -iGs \left[\log \left(\frac{\rho}{2Gm^{(2)}}\right)^2
				   -\frac{\pi}{4} \left(\frac{2Gm^{(2)}}{\rho} \right)
				         \right] \right\}~~.\nonumber\\
\end{eqnarray}
Here $k_{-}=\left( \tilde{k}+m^{(1)2} \right)/k_{+}$,
and $k_{+}$ and $k_{-}$ are conjugate momenta of $u$ and $v$, respectively.
The integral region of $\rho$ is from 0 to $\infty$,
but the small region of $\rho$ contributes little
because the amplitude oscillates fast enough.
Therefore only the large region of $\rho$, is to be considered.
For $Gm^{(2)}/\rho \ll 1$, the $A \left( k_+, \tilde{k} \right)$ is :
\begin{eqnarray}
&&A \left( k_+, \tilde{k} \right)~\sim~
        \delta \left( k_+ -p_+^{(1)} \right) \nonumber\\
		&&~~\times \frac{1}{\left(2 \pi \right)^2} \int d^2 \tilde{x}
		    \exp^{i\tilde{k} \tilde{x} }
			    \left(\frac{\rho}{2Gm^{(2)}}\right)^{-2iGs}
				   \left\{1+i\frac{\pi}{4} Gs
				       \left(\frac{2Gm^{(2)}}{\rho} \right)\right\}~~. \nonumber\\
\end{eqnarray}
Using the following integral formula:
\begin{eqnarray}
\int d^2 \tilde{x} \exp^{i\tilde{k} \tilde{x} } \left( \rho^2 \right)^{\lambda}
   ~=~4^{\lambda +1} \pi
     \frac{\Gamma \left( \lambda+1 \right)}{\Gamma \left( -\lambda \right)}
         \left( k^2 \right)^{-\lambda-1}~~,
\end{eqnarray}
the S matrix becomes:
\begin{eqnarray}
&&{}_{out}\left\langle \mbox{\boldmath$k$}^{(1)}
|\mbox{\boldmath$p$}^{(1)} \right\rangle_{in} \nonumber\\
&&= \frac{k_{+}}{4 \pi k_0} \delta \left(k_+-p_+ \right)
   \left( 2Gm^{(2)} \right)^{2iGs} \left(\frac{4}{-t} \right)^{1-iGs}
       f \left(s,t \right)~~,
\end{eqnarray}
where
\begin{eqnarray}
f \left( s,t \right)~=~
     \frac{\Gamma \left(1-iGs \right)}{\Gamma \left( iGs \right)}
	     +i \frac{\pi}{4} Gs \left(2Gm^{(2)} \right)
		     \left( \frac{-t}{4} \right)^{1/2}
			      \frac{\Gamma \left( \frac{1}{2}-iGs \right)}
				        {\Gamma \left( \frac{1}{2}+iGs \right)}~~.
\end{eqnarray}
We put the Mandelstam variable
$t=-q^2=-\left[ \left(k^{(1)} -p^{(1)} \right)^2 \right]$.
The first term is that given by 't Hooft \cite{'t Hooft87},
and  the second term is the correction to it.

The scattering amplitude is now given by:
\begin{eqnarray}
U \left( s,t \right)~=~\frac{1}{4 \pi} \left( 2Gm^{(2)} \right)^{2iGs}
    \left(\frac{4}{-t} \right)^{1-iGs}
       f \left(s,t \right)~~.
\end{eqnarray}
{}From this, the cross section is derived:
\begin{eqnarray}
\sigma \left( \tilde{p}^{(1)} \rightarrow \tilde{k}^{(1)} \right) d^2
\tilde{k}~=~
\frac{4}{t^2} \left| f \left( s,t \right) \right|^2 d^2\tilde{k}~~,
\end{eqnarray}
where
\begin{eqnarray}
\left| f \left( s,t \right) \right|^2~&=&~G^2 s^2
   \left\{ 1-\frac{\pi}{2} Gm^{(2)} \sqrt{-t} \cos 2\theta
      +\frac{\pi^2}{16} \left( Gm^{(2)} \right)^2
	        \left( -t \right) \right\}~~,\nonumber\\
\theta~&=&~\sum^{\infty}_{n=0} \left( \tan^{-1} \frac{Gs}{n}
   -\tan^{-1} \frac{Gs}{n+\frac{1}{2}} \right)~~.
\end{eqnarray}
It has a very simple form!
For $Gs \gg 1$, Eq.(24) is further simplified as:
\begin{eqnarray}
\sigma \left( \tilde{p}^{(1)} \rightarrow \tilde{k}^{(1)} \right) d^2 \tilde{k}
{\longrightarrow \atop Gs \gg 1}
\frac{4G^2 s^2}{t^2} \left( 1-\frac{\pi}{4} Gm^{(2)} \sqrt{-t} \right)^2
    d^2\tilde{k}~~.
\end{eqnarray}

The first and the third term
of Eq.(25) represent the one graviton exchange and the two gravitons exchange,
respectively, whereas the second term is the interference one.

Finally, consider the case of the particle (2) as a massive--charged particles
($m^{(2)} \sim M_{Planck}$ and $\sqrt{G}m^{(2)} \sim e$).
The particle (1) is same as  above ($m^{(2)} \gg m^{(1)}$).
The cross section is given by:
\begin{eqnarray}
\sigma \left( \tilde{p}^{(1)} \rightarrow \tilde{k}^{(1)} \right) d^2 \tilde{k}
{\longrightarrow \atop Gs \gg 1}
\frac{4G^2 s^2}{t^2} \left( 1-\frac{\pi}{16} Gm^{(2)} \sqrt{-t} \right)^2
    d^2\tilde{k}~~,
\end{eqnarray}
which has the same form as Eq.(26), apart from a numerical factor.

The calculation above is performed only perturbatively in $1/\rho$ expansion.
It is desirable to obtain the full result without approximations.
In that case the nonlinear effects of gravitation would manifest,
and would shed a light to the theory of quantum gravity.

\end{document}